\documentclass[aps,prl,reprint,groupedaddress,footinbib]{revtex4-2}

\usepackage[utf8]{inputenc}
\usepackage{latexsym, graphicx} 
\usepackage{amsmath,amsfonts,amssymb}
\usepackage{cancel}
\usepackage{mathrsfs,dsfont}
\usepackage{tensor}
\usepackage{xcolor}
\usepackage{caption,subcaption}
\usepackage[shortlabels]{enumitem}
\usepackage{hyperref}
\hypersetup{
    colorlinks,%
    citecolor=blue,%
    filecolor=blue,%
    linkcolor=blue,%
   	urlcolor=blue,
   	linktoc=page
}

\begin{document}
\title{Massless contraction of massive particles in the infinite boost limit}
\author{Kevin Nguyen}
\email{kevin.nguyen2@ulb.be}

\affiliation{\vspace{0.2cm} Universit\'e Libre de Bruxelles and International Solvay Institutes, ULB-Campus Plaine
CP231, 1050 Brussels, Belgium}

\begin{abstract} 
Massless particles, defined through Wigner's induced representations of the Poincar\'e group, are obtained as a limit of infinitely boosted massive particles wherein the massive little group undergoes an Inönü--Wigner contraction to the massless one. It is demonstrated that the spin-$s$ multiplet of a massive representation re-organizes into the direct sum of massless particles with helicities in the range $[-s,s]$. This group theoretic result is the Hilbert space counterpart of the Stückelberg formulation of massive fields of spin $s$ in the high-energy limit, where the Stückelberg fields become independent massless fields of spins $s'\leq s$ in integer steps. It therefore also provides the kinematic setting of the Goldstone equivalence theorem.
\end{abstract}
\maketitle
  
\section{Introduction}
The purpose of this letter is to describe in detail how massless particles result from a high-energy limit of massive particles. The definition of `particle' which we adopt here is that introduced by Wigner \cite{Wigner:1939cj}, i.e., a unitary irreducible representation of the proper orthochronous inhomogeneous Lorentz group $\operatorname{ISO}(1,3)^\uparrow$, also known as Poincaré group. This notion of particle is fundamental in the theory of particle scattering, as described in many physics textbooks \cite{Weinberg:1995mt,Martin:1970hmp,Hebbar:2020ukp}. It is intuitively clear that amplitudes describing the scattering of massless particles should result from a suitable high-energy limit of amplitudes describing the scattering of massive particles. This has been described in recent works using spinor-helicity variables \cite{Arkani-Hamed:2017jhn,Chiodaroli:2022ssi,Diaz-Cruz:2026jdi}. In the present work, we wish to address the same question directly at the level of the one-particle Hilbert space, rather than at the level of scattering amplitudes. 

The high-energy limit producing a massless representation from a massive one is properly understood in terms of an Inönü--Wigner contraction \cite{Inonu:1953sp,Saletan:1961iqo,Mickelsson:1972fh}. Indeed, Han, Kim, Son, Noz and Wigner had shown some years ago that infinite boost yields a contraction of the massive little group $\operatorname{SO}(3)$ to the massless little group $\operatorname{ISO}(2)$ \cite{Han:1983by,Han:1984xp,Kim:1989wt,Kim:2001wd}. Building upon this observation, we will describe in detail the contraction of a massive spin-$s$ representation in the infinite boost limit. In particular, we will show that the resulting massless representation is fully reducible and decomposes into the direct sum of massless particles with helicity ranging from $-s$ to $s$ in integer steps,
\begin{equation}
\label{direct sum}
\operatorname{V}_{m,s} \quad \stackrel{\text{high-energy}}{\longrightarrow} \quad \bigoplus_{h=-s}^s \operatorname{V}_{0,h}\,.
\end{equation}
This result has already been obtained in two different ways: first, in a limit where the mass parameter itself is taken to zero  \cite{Robinson:1962zeromass,Coester:1963zeromass,Korff:1964zeromass,McKerrell:1965canonical}; second, from the $\operatorname{SO}(3) \downarrow \operatorname{SO}(2)$ branching rules that underlie the contraction of the massive little group \cite{Bekaert:2006py}. Our treatment differs from these in that the mass parameter is kept fixed: the limit is taken along the orbit of a fixed massive representation, so that the contraction of the little group is a consequence of the construction rather than an input. This makes it possible to identify what carries the contraction, namely a single dilation of the momenta which commutes with the little group and induces a helicity-independent rescaling of the inner product.

As should be expected, the contraction \eqref{direct sum} supports the results obtained in the high-energy limit of scattering amplitudes put forward in \cite{Arkani-Hamed:2017jhn,Chiodaroli:2022ssi,Diaz-Cruz:2026jdi}. More generally, we make the observation that it mirrors the Stückelberg formulation of massive fields in the high-energy limit, where a massive field of spin $s$ is traded for a set of massless gauge fields of spins $s' \leq s$ \cite{Stueckelberg:1957zz,deRham:2014zqa}. As a byproduct, it therefore also provides the kinematic setting for the Goldstone equivalence theorem \cite{Cornwall:1974km,Vayonakis:1976vz,Lee:1977eg,Chanowitz:1985hj,Veltman:1989ud,Horejsi:1995jj,Graf:2022rco,Wulzer:2013mza,Cuomo:2019siu} stating that the longitudinal modes of a massive field are described, at energies much larger than its mass, by the corresponding would-be Goldstone bosons/Stückelberg fields. 

This letter is organized as follows. We start by recalling Wigner's construction of the massive particles in terms of induced representations. We then describe the contraction of the massive little group to the massless one that results from an infinite boost, following the seminal work done in \cite{Han:1983by,Han:1984xp,Kim:1989wt,Kim:2001wd}. In the next section, we provide a detailed description of the resulting contraction of massive to massless representations in their standard reference frames. In particular, we show that the resulting massless representations are fully reducible and we derive the decomposition stated in \eqref{direct sum}. We then complete the description and generalize to arbitrary momentum frames, by implementing the induction procedure of Wigner. We end by discussing the relation of this group theoretic result to the Stückelberg formulation of massive field theories and to the Goldstone equivalence theorem.

\section{Massive representations}
\label{sec 2}
Let us first recall the construction of massive unitary irreducible representation (UIR) of the Poincaré group, following Wigner's method of induced representations \cite{Wigner:1939cj,Weinberg:1995mt,Martin:1970hmp,Hebbar:2020ukp}. We adopt the following representation of the $\mathfrak{iso}(1,3)$ algebra,
\begin{align}
\left[L_{\mu\nu},L_{\rho\sigma}\right]&=i\left(\eta_{\mu \rho} L_{\nu \sigma}+\eta_{\nu\sigma}L_{\mu\rho}-\eta_{\mu\sigma} L_{\nu \rho}-\eta_{\nu\rho} L_{\mu\sigma} \right)\,, \nonumber\\
\left[L_{\mu\nu},P_\rho \right]&=i \left(\eta_{\mu\rho}P_\nu-\eta_{\nu\rho}P_\mu \right)\,,\\
\left[P_\mu,P_\nu\right]&=0\,,\nonumber
\end{align}
where $\eta_{\mu\nu}=\text{diag}(-,+,+,+)$. With these conventions, unitarity of a representation is equivalent to hermiticity of the Lorentz and momentum generators $L_{\mu\nu}$ and $P_{\mu}$.

A massive UIR of the Poincaré group, labeled by a mass $m>0$ and a spin $s \in \frac{1}{2} \mathbb{Z}$, and denoted $\operatorname{V}_{m,s}$ here, is constructed from the reference momentum
\begin{equation}
\label{rest momentum}
k^\mu= m\, \delta^\mu_0\,,
\end{equation}
whose orbit under the Lorentz group is the mass-shell $p^2=-m^2$. One associates a spin-$s$ representation to the little group $\operatorname{SO}(3)$ consisting of rotations leaving $k^\mu$ invariant. A spin-$s$ representation of the rotation group can be described in terms of eigenstates of $\vec L^2\equiv \frac{1}{2} L_{ij} L^{ij}$ and $L_{12}$,
\begin{equation}
\label{spin states}
\begin{split}
\vec L^2 |m, s h\rangle&=s (s+1) |m, s h\rangle\,,\\
L_{12} |m, s h\rangle&=h |m, s h\rangle\,,
\end{split}
\end{equation}
with \textit{helicity} taking values $h=-s,-s+1,...,s$. A massive spin-$s$ multiplet therefore contains $(2s+1)$ helicity states. The next step is to construct states with generic momentum $p^\mu$ on the same mass-shell, which we parametrize as
\begin{equation}
\label{generic momentum}
\begin{split}
p^0&=\sqrt{m^2+\mathbf{p}^2}\,,\\
\vec p&=\mathbf{p}\, (\sin \theta \cos \varphi\,, \sin \theta \sin \varphi\,, \cos \theta)\,.
\end{split}
\end{equation}
This generic momentum is related to the reference momentum \eqref{rest momentum} by a boost matrix,
\begin{equation}
\label{boost matrix}
p^\mu=L(p)\indices{^\mu_\nu}\, k^\nu\,, 
\end{equation}
whose explicit expression may be found for instance in \cite{Weinberg:1995mt,Iacobacci:2024laa}. It is always possible to decompose the corresponding boost element as \cite{Weinberg:1995mt,Martin:1970hmp,Hebbar:2020ukp}
\begin{equation}
\label{boost element}
L(p)=R(\theta, \varphi) B(\mathbf{p})\,, 
\end{equation}
where $B(\mathbf{p})$ is a pure boost along the third axis and $R(\theta,\varphi)$ is a spatial rotation. The boost element is given by
\begin{equation}
\label{B(eta)}
B(\mathbf{p})\equiv B(\eta)=e^{i\eta L_{03}}\,, 
\end{equation}
written in terms of the rapidity $\eta$ satisfying
\begin{equation}
\cosh \eta=\frac{p^0}{m}\,, \qquad \sinh \eta=\frac{\mathbf{p}}{m}\,.
\end{equation}
The rotation element, following the convention of Jacob and Wick \cite{Jacob:1959at}, is given by 
\begin{equation}
R(\theta,\varphi)=e^{-i\varphi L_{12}} e^{i \theta L_{13} } e^{i \varphi L_{12}}\,.
\end{equation}
A generic momentum state $|p,h\rangle$ in the \textit{helicity basis} is then obtained by application of the boost element \eqref{boost element} to the states \eqref{spin states} \cite{Martin:1970hmp,Hebbar:2020ukp},
\begin{equation}
\label{generic momentum state}
|p, h \rangle \equiv U(L(p)) |m, sh \rangle\,,
\end{equation}
where the labels $m,s$ are left implicit.
It is easily shown that these states are still momentum eigenstates, namely
\begin{equation}
P_\mu |p, h \rangle = U(L(p))L(p)\indices{_\mu^\nu} P_\nu |m, sh \rangle=p_\mu\, |p, h \rangle\,.
\end{equation}
Acting with an element of the Lorentz group $\Lambda \in \operatorname{SO}(1,3)$, a standard argument yields the transformation rule \cite{Weinberg:1995mt}
\begin{equation}
U(\Lambda) |p,h\rangle = \sum_{h'} [W(\Lambda,p)^{-1}]^{(s)*}_{hh'} |\Lambda p, h'\rangle\,,
\end{equation}
written in terms of the Wigner rotation $W(\Lambda,p) \in \operatorname{SO}(3)$, 
\begin{equation}
W(\Lambda, p)=L^{-1}(\Lambda p) \Lambda L(p)\,,
\end{equation}
with $[W]^{(s)}$ the corresponding spin-$s$ matrix representation (Wigner $D$-matrix). 

We now have at our disposal everything we need to discuss the Inönü--Wigner contraction of the massive representations to massless ones in the limit of infinite boost.

\section{Contraction of the little group}
\label{sec 3}
We start by recalling the crucial observation made in \cite{Han:1983by,Han:1984xp,Kim:1989wt,Kim:2001wd}, that the massive little group $\operatorname{SO}(3)$ contracts to the massless little group $\operatorname{ISO}(2)$ in the limit of infinite boost. To show this, we start again from the rest frame of the massive particle characterized by \eqref{rest momentum}-\eqref{spin states}. We then boost the particle momentum along the third axis, by application of the group element $B(\eta)$ as defined in \eqref{B(eta)}. By the definition \eqref{generic momentum state}, the resulting state is
\begin{equation}
\label{boosted particle}
|p(\eta) , h \rangle \equiv B(\eta) |m, s h\rangle\,, 
\end{equation}
with momentum
\begin{equation}
\label{boosted momentum}
p^\mu(\eta)= (m \cosh \eta, 0, 0, m \sinh \eta)\,.
\end{equation}
Under this boost, Lorentz generators themselves transform by the inner automorphism
\begin{equation}
\nonumber
L_{\mu\nu}(\eta)\equiv B(\eta) L_{\mu\nu} B(\eta)^{-1}=[B(\eta)]\indices{^\rho_\mu} [B(\eta)]\indices{^\sigma_\nu} L_{\rho \sigma}\,.
\end{equation}
In particular, the generators of the massive little group $\operatorname{SO}(3)$ transform to 
\begin{equation}
\label{boosted rotations}
\begin{split}
L_{12}(\eta)&=L_{12}\,,\\
L_{13}(\eta)&=\cosh \eta\, L_{13}-\sinh \eta\, L_{01}\,,\\
L_{23}(\eta)&=\cosh \eta\, L_{23}-\sinh \eta\, L_{02}\,.
\end{split}
\end{equation}

In the infinite boost limit $\eta \to \infty$, the momentum $p^\mu(\eta)$ becomes effectively null. More precisely, we can introduce the rescaled momentum $q^\mu(\eta)$, 
\begin{equation}
\label{rescaled momentum}
q^\mu(\eta)\equiv 2 e^{-\eta} p^\mu(\eta)\,, 
\end{equation}
which is well-defined and null in that limit,
\begin{equation}
\label{rescaled momentum bis}
q^\mu\equiv \lim_{\eta \to \infty} q^\mu(\eta)= (m,0,0,m)\,.
\end{equation}
Modulo normalization, this is precisely the standard reference momentum used to construct massless Poincaré particles through Wigner's method of induced representations \cite{Weinberg:1995mt}. At the level of the algebra generators, \eqref{rescaled momentum}-\eqref{rescaled momentum bis} are achieved by introducing the rescaled momentum generators
\begin{equation}
\label{rescaled momentum generators}
Q_\mu(\eta) \equiv 2 e^{-\eta} P_\mu\,, \qquad Q_\mu\equiv \lim_{\eta \to \infty} Q_\mu(\eta)\,.
\end{equation}
Similarly, we can rescale the boosted rotation generators,
\begin{equation}
\begin{split}
N_1(\eta)&\equiv 2 e^{-\eta} L_{13}(\eta)\,,\\
N_2(\eta)&\equiv 2 e^{-\eta} L_{23}(\eta)\,, 
\end{split}
\end{equation}
and observe that these precisely reduce to the `translation' generators of the little group $\operatorname{ISO}(2)$ leaving the null momentum \eqref{rescaled momentum bis} invariant,
\begin{equation}
\label{N generators}
\begin{split}
N_1\equiv \lim_{\eta \to \infty} N_1(\eta) = L_{13}-L_{01}\,,\\
N_2\equiv \lim_{\eta \to \infty} N_2(\eta) = L_{23}-L_{02}\,.
\end{split}
\end{equation}
Together with $L_{12}$, they satisfy the $\mathfrak{iso}(2)$ commutation relations
\begin{equation}
[L_{12},N_1]=iN_2, \,\, [L_{12},N_2]=-i N_1,\,\, [N_1,N_2]=0\,.
\end{equation}

The above manipulations suggest that massless Poincaré representations are obtained from an Inönü--Wigner contraction of the massive Poincaré representations in the infinite boost limit $\eta \to \infty$. The momentum rescaling $P_\mu \mapsto \lambda P_\mu$ used in \eqref{rescaled momentum}--\eqref{rescaled momentum generators}, with $\lambda=2e^{-\eta}$, is an automorphism of
$\mathfrak{iso}(1,3)$ and maps $\mathcal{V}_{m,s}$ to
$\mathcal{V}_{\lambda m,s}$. Composed with the boost $B(\eta)$ which is inner and leaves the mass unchanged, we are therefore considering a family
of representations $\mathcal{V}_{m_\eta,s}$ with rescaled mass
$m_\eta \equiv 2 e^{-\eta} m \to 0$ and rescaled energy approaching $m$. This is the
precise sense in which the infinite boost limit is equivalent to the
massless limits previously studied in \cite{Robinson:1962zeromass,Coester:1963zeromass,Korff:1964zeromass,McKerrell:1965canonical}. 

Note that the full algebra $\mathfrak{iso}(1,3)$ does not undergo an actual contraction, by contrast to the little algebra, since the resulting generators $\{Q_\mu, N_1, N_2, L_{12}, L_{0i} \}$ still close to an $\mathfrak{iso}(1,3)$ algebra. What is less trivial, on the other hand, is the contraction of the massive irreps to massless representations obtained through a rescaling of the Hilbert space inner product. We detail this phenomenon in the following section.

\section{Massless helicity states}
\label{sec 4}
Thus far, we understand how to obtain the reference momentum \eqref{rescaled momentum bis} of a massless particle from the momentum of a massive particle boosted from its rest frame. We also understand how the massive little group contracts to the massless little group. We are left to understand the re-organization of the $(2s+1)$ massive spin states into massless helicity states. First, we formally define the reference states in the contracted representation as
\begin{equation}
|q, h\rangle \equiv \lim_{\eta \to \infty} |p(\eta),h\rangle\,. 
\end{equation}
More precisely, we are considering the convergence of matrix elements with respect to the rescaled sesquilinear form $\langle \cdot | \cdot \rangle_\eta \equiv \frac{1}{4} e^{2\eta} \langle \cdot | \cdot \rangle$ in the limit $\eta \to \infty$ .
Since $L_{12}$ commutes with the boost element $B(\eta)$, these states still satisfy
\begin{equation}
L_{12} |q, h \rangle = h|q, h \rangle\,.  
\end{equation}
This agrees with the definition of helicity $h$ for a massless particle, which in this case takes values $h=-s,-s+1,...,s$. Besides this, massless particles are also characterized by a trivial action of the little group generators $N_1, N_2$. Let us check that this is indeed the case. As an intermediate step, using $|p(\eta),h\rangle =B(\eta)|m,sh\rangle$, we compute
\begin{equation}
\label{eq3.14}
\begin{split}
L_{13}(\eta) |p(\eta), h \rangle
&=B(\eta) L_{13} |m,sh\rangle\\
&=\sum_{h'} [L_{13}]^{(s)}_{h'h}|p(\eta), h' \rangle\,,
\end{split}
\end{equation}
which is well-defined and finite in the limit $\eta \to \infty$, by contrast to what could have been naively expected from \eqref{boosted rotations}. This implies that $N_1$ acts trivially in the contracted representation, 
\begin{equation}
N_1 |q,h\rangle \equiv \lim_{\eta \to \infty} 2 e^{-\eta} L_{13}(\eta) |p(\eta),h\rangle=0\,.
\end{equation}
By the same argument, $N_2$ also acts trivially. Hence, the representation resulting from the Inönü--Wigner contraction fulfills all the requirements that characterize the massless particle representations of the little group $\operatorname{ISO}(2)$. Furthermore, it can be decomposed into a direct sum of massless UIRs, 
\begin{equation}
\label{direct sum bis}
\operatorname{V}_{m,s} \quad \stackrel{\eta \to \infty}{\longrightarrow} \quad \bigoplus_{h=-s}^s \operatorname{V}_{0,h}=\bigoplus_{s'=s-\lfloor s\rfloor }^s \operatorname{V}^{(\mathcal{P})}_{0,s'}\,.
\end{equation}
Each massless UIR $\operatorname{V}_{0,h}$ contains a single helicity state, transforming into itself under the proper orthochronous Lorentz group. Opposite helicities $h=\pm s'$ (for $s' \neq 0$) transform into each other under reflection symmetry $\mathcal{P}$ and are therefore paired up inside $\operatorname{V}^{(\mathcal{P})}_{0,s'}$. Together, they make up the $(2s+1)$ spin multiplet we started from. Note that $s-\lfloor s\rfloor$ equals $0$ for integer $s$ and $1/2$ for half-integer $s$.

\section{Generic momentum frame}
To complete the analysis, we should also implement the contraction on states with generic momentum $p^\mu$ on the mass-shell $p^2=-m^2$. We recall the definition of a generic massive particle state given in \eqref{generic momentum state}, namely
\begin{equation}
\label{generic momentum state bis}
\begin{split}
|p(\eta,\theta,\varphi), h \rangle &\equiv R(\theta,\varphi) B(\eta) |m, sh \rangle\\
&=R(\theta,\varphi) |p(\eta), h\rangle\,.
\end{split}
\end{equation}
We see that they are obtained from the boosted particles \eqref{boosted particle} of the previous section through the rotation $R(\theta,\varphi)$. Implementing the infinite boost limit $\eta \to \infty$, which commutes with the rotation, we obtain 
\begin{equation}
|q(\theta,\varphi), h \rangle \equiv \lim_{\eta \to \infty} |p(\eta,\theta,\varphi), h \rangle=R(\theta,\varphi)  |q, h \rangle\,.
\end{equation}
The resulting relation exactly coincides with the standard definition \cite{Weinberg:1995mt} of a massless particle of generic null momentum 
\begin{equation}
\label{q with fixed m}
q^\mu(\theta,\varphi)=m\, (1,\sin \theta \cos \varphi\,, \sin \theta \sin \varphi\,, \cos \theta)\,,
\end{equation}
where the original mass parameter $m$ now appears in place of the energy. The same conclusion is reached by evaluating the action of the rescaled momentum $Q_\mu(\eta)$ before taking the infinite boost limit,
\begin{equation}
\begin{split}
Q_\mu |q(\theta,\varphi), h \rangle &=\lim_{\eta \to \infty} Q_{\mu}(\eta) |p(\eta,\theta,\varphi), h \rangle\\
&=q_\mu(\theta,\varphi) |q(\theta,\varphi),h \rangle\,.
\end{split}
\end{equation}
In order to really cover the full orbit $q^2=0$, we can boost the massive particle \eqref{generic momentum state bis} in a way that implements the replacement $m \mapsto \omega$ in \eqref{q with fixed m},
\begin{equation}
\label{generic massless state}
\begin{split}
|q(\omega,\theta,\varphi), h \rangle &\equiv \lim_{\eta \to \infty} |p(\eta+\zeta,\theta,\varphi), h \rangle\\
&= R(\theta,\varphi) B(\zeta) |q,  h \rangle\,, 
\end{split}
\end{equation}
with additional boost parameter $\zeta=\ln \frac{\omega}{m}$. 
Because \eqref{generic massless state} now exactly coincides with the definition of a massless particle state according to Wigner's method \cite{Weinberg:1995mt}, its behavior under the action of the Poincaré group resulting from the contraction is the standard one. 

Finally, we can consider the limit of the integral measure on the mass-shell $p^2=-m^2$. Because it is Lorentz-invariant, it can easily be written in coordinates $(\mathbf{p},\theta,\varphi)$ as
\begin{equation}
\nonumber
[d^3p]= \frac{1}{(2\pi)^3}\frac{d^3 \vec{p}}{2\sqrt{m^2+\mathbf{p}^2}}=\frac{\sin \theta\, d\theta\, d\varphi}{(2\pi)^3}\frac{\mathbf{p}^2\, d \mathbf{p}}{2\sqrt{m^2+\mathbf{p}^2}}\,,   
\end{equation}
with $\mathbf{p}=m \sinh (\eta+\zeta)$. In the limit $\eta \to \infty$, one recovers the Lorentz-invariant measure on the null cone $q^2=0$ up to a suitable rescaling, i.e.,
\begin{equation}
\begin{split}
[d^3q]&= \lim_{\eta \to \infty} 4e^{-2\eta} [d^3p]\\
&=\frac{\sin \theta\, d\theta\, d\varphi}{(2\pi)^3}\frac{\omega\, d\omega}{2}=\frac{1}{(2\pi)^3}\frac{d^3 \vec{q}}{2|\vec q\,|}\,.
\end{split}
\end{equation}
A dual statement applies to the inner product on the one-particle Hilbert spaces, namely
\begin{equation}
\label{rescaled norm}
\begin{split}
\langle q_1,h_1 |q_2,h_2\rangle &= \lim_{\eta \to \infty} \frac{e^{2\eta}}{4} \langle p_1,h_1 |p_2,h_2\rangle\\
&=2 (2\pi)^3 |\vec q\,| \delta^{(3)}(\vec q_1-\vec q_2) \delta_{h_1,h_2}\,.
\end{split}
\end{equation}
The contraction thus reproduces the massless one-particle Hilbert space in full, provided states, measure and inner product are rescaled consistently.

\section{Discussion}
\label{sec 5}

In this letter, we provided a concise but detailed description of the massless limit of massive particle representations of the Poincar\'e group, implemented through an infinite boost limit and building upon earlier the known little group contraction \cite{Han:1983by,Han:1984xp,Kim:1989wt,Kim:2001wd}. Working with Wigner's induced representations throughout, we followed the contraction on states of generic momentum, on the invariant measure and on the inner product, and
found that it is carried in its entirety by a single helicity-independent rescaling. This complements recent analyses performed directly at the level of scattering amplitudes
\cite{Arkani-Hamed:2017jhn,Chiodaroli:2022ssi,Diaz-Cruz:2026jdi}, and provides the underlying contraction of the one-particle Hilbert space in the high-energy limit.

The main physical result of this contraction is that the $(2s+1)$ massive helicity states of the irreducible representation become helicity states for massless particles with integer spins $s'=0,...,s$ or half-integer spins $s'=1/2,...,s$. This decomposition is encapsulated by the formula \eqref{direct sum bis}. Interestingly, it can be seen to mirror the St\"uckelberg formulation of massive field theories in the high-energy limit.  As a concrete
example, let us consider the Fierz--Pauli field $h_{\mu\nu}$ describing massive
spin-2 degrees of freedom. The Stückelberg substitution \cite{deRham:2014zqa}
\begin{equation}
\label{Stuckelberg}
    h_{\mu \nu} \; \longrightarrow \; h_{\mu \nu}
    + \frac{1}{m} \partial_{(\mu} A_{\nu)}
    + \frac{1}{m^2} \partial_\mu \partial_\nu \pi \, ,
\end{equation}
restores the gauge symmetries 
\begin{equation}
\nonumber
\delta h_{\mu\nu} = \partial_{(\mu} \xi_{\nu)}\,, \quad \delta A_\mu =-m\, \xi_\mu +\partial_\mu \lambda\,, \quad 
\delta \pi = - m\, \lambda\,,
\end{equation}
which would otherwise be broken by the mass term. Hence $A_\mu$ and $\pi$ are would-be Goldstone modes associated to explicit breaking of gauge symmetries. At energies $E \gg m$, the five states of the massive multiplet are then carried by $h_{\mu\nu}$, $A_\mu$
and $\pi$, respectively propagating massless particles of helicities $\pm 2$, $\pm 1$ and $0$.
This corresponds precisely to the pattern we derived in the infinite boost limit \eqref{direct sum bis}  for $s = 2$.

In relation to the previous point, the contraction \eqref{direct sum bis} is also the one-particle Hilbert space counterpart of the Goldstone equivalence theorem \cite{Cornwall:1974km,Vayonakis:1976vz,Lee:1977eg,Chanowitz:1985hj,Veltman:1989ud,Horejsi:1995jj,Graf:2022rco} stating that amplitudes involving longitudinal polarizations of massive bosons of energy $E \gg m$ coincide, up to corrections of order $O(m/E)$, with the
amplitudes of the corresponding Goldstone/Stückelberg particles. In particular, the longitudinal polarization vector associated with a massive vector field of boosted momentum \eqref{boosted momentum} may be written 
\begin{equation}
\label{polarization eta}
\begin{split}
\epsilon^\mu_L(p(\eta))=(\sinh \eta,0,0, \cosh \eta)=\frac{p^\mu(\eta)}{m}-e^{-\eta} n^\mu\,, 
\end{split}
\end{equation}
with $n^\mu=(1,0,0,-1)$ a fixed null vector identified with the boost axis. This is a concrete incarnation of the relation used to establish the equivalence theorem \cite{Chanowitz:1985hj,Veltman:1989ud}, where contraction of $p^\mu/m$ with the massive amplitude $\mathcal{M}_{\mu...}(p,...)$ is shown to equal the amplitude involving a Goldstone boson in place of the longitudinal vector, and with $e^{-\eta}$ controlling the $O(m/E)$ deviations. Going one step further and rewriting \eqref{polarization eta} covariantly,
\begin{equation}
\begin{split}
\epsilon^\mu_L(p)=\frac{p^\mu}{m}+ \frac{m\, n^\mu}{n \cdot p}\,,
\end{split}
\end{equation}
we recognize the form of the longitudinal polarization employed in the `equivalent gauge' formulation of massive gauge theories \cite{Wulzer:2013mza,Cuomo:2019siu}, where the first term is traded for the Goldstone polarization through Ward identities, and the second term still vanishes like $\sim m/E$. In the standard covariant formulation, the
high-energy limit is obstructed by the apparent growth $\epsilon^\mu_L \sim E/m$, or
$\epsilon^{\mu\nu}_L \sim (E/m)^2$ for the longitudinal mode of $h^{\mu\nu}$,
which is eventually avoided after a rearrangement of the states or of the Feynman
rules \cite{Wulzer:2013mza,Cuomo:2019siu}. In the Hilbert space contraction presented here, no such apparent obstruction arises at
any stage: $B(\eta)$ is unitary, so the $(2s+1)$ states retain equal norms at
every rapidity $\eta$, and the only $\eta$-dependence is the overall factor $e^{2\eta}/4$ used to rescale
\eqref{rescaled norm}, whose origin is merely the conversion between the
massive and massless momenta in \eqref{rescaled momentum}.
What grows anomalously is thus the normalization of the polarization tensors that intertwine the covariant field description with the Wigner representation.

As a final remark, the infinite boost limit described in this paper will be shown to play an important role in the flat limit of the AdS/CFT correspondence \cite{Nguyen:appear}, which constitutes the original motivation for carrying out this work, as part of the author's effort \cite{Nguyen:2023vfz,Nguyen:2023miw,Have:2024dff,Nguyen:2025sqk,Agrawal:2025bsy} to develop an understanding of Carrollian holography \cite{Bagchi:2025vri,Nguyen:2025zhg,Ruzziconi:2026bix} from first principles.

\section{Acknowledgments}
This work was supported by a postdoctoral research fellowship of the F.R.S.-FNRS (Belgium).

\bibliography{bibl}
\bibliographystyle{JHEP}  

\end{document}